%% file: main.tex
\documentclass[acmsmall,screen]{acmart}
\renewcommand\footnotetextcopyrightpermission[1]{} 
\usepackage{multirow,makecell}
\usepackage[most]{tcolorbox}
\usepackage{color,xcolor}
\usepackage{listings,amsfonts}
\usepackage{caption}
\usepackage{subcaption}
\usepackage{threeparttable}
\usepackage{bbding}
\usepackage{graphicx}
\usepackage{booktabs} 
\usepackage{longtable}
\usepackage{enumitem}
\usepackage[table]{xcolor}
\usepackage{makecell}
\usepackage{array} 
\usepackage{longtable,booktabs}

\definecolor{heatmap0}{HTML}{fafdcf}  
\definecolor{heatmap33}{HTML}{e8f6b1}
\definecolor{heatmap66}{HTML}{cfecb3}
\definecolor{heatmap90}{HTML}{c1e7b4}  
\definecolor{heatmap100}{HTML}{9cd8b8} 

\definecolor{customcite}{HTML}{b83b5e}
\definecolor{customlink}{HTML}{07689f}
\definecolor{customurl}{HTML}{1fab89}
\AtEndPreamble{
\usepackage{hyperref}

    \hypersetup{
      colorlinks = true,
      linkcolor = customlink,
      anchorcolor = purple,
      citecolor = customcite,
      filecolor = purple,
      urlcolor = customurl
    }
}

\definecolor{jsonstring}{HTML}{D19A66}
\definecolor{jsonnumber}{HTML}{3B78E7}
\definecolor{jsoncomment}{HTML}{50A14F}
\definecolor{jsonbg}{HTML}{F8F8F8}

\usepackage{listings}
\lstdefinestyle{customjson}{
  backgroundcolor=\color{jsonbg},
  basicstyle=\ttfamily\scriptsize,
  frame=single,
  breaklines=true,
  showstringspaces=false,
  tabsize=2,
  numbers=left,
  numberstyle=\color{jsonnumber}\scriptsize\ttfamily,
  commentstyle=\color{jsoncomment},
  morecomment=[l]{//},
  stringstyle=\color{jsonstring},
  keywordstyle=\color{black},
}

\definecolor{customcell}{HTML}{E6F2FF}

\def\BibTeX{{\rm B\kern-.05em{\sc i\kern-.025em b}\kern-.08em
    T\kern-.1667em\lower.7ex\hbox{E}\kern-.125emX}}

\newtcolorbox{takeawaybox}{
  colback=white,
  colframe=black,
  boxrule=0.6pt,
  arc=2.5pt,
  left=2pt,
  right=2pt,
  top=2pt,
  bottom=2pt,
  before skip=4pt,
  after skip=4pt
}

\begin{document}

\title{SMCP: Secure Model Context Protocol}

\author[X Hou]{Xinyi Hou}
\email{xinyihou@hust.edu.cn}
\authornote{Xinyi Hou and Shenao Wang contributed equally to this work.}
\affiliation{%
  \institution{Huazhong University of Science and Technology}
  \city{Wuhan}           
  \country{China}
}

\author[S Wang]{Shenao Wang}
\email{shenaowang@hust.edu.cn}
\authornotemark[1]
\affiliation{%
  \institution{Huazhong University of Science and Technology}
  \city{Wuhan}           
  \country{China}
}

\author[Y Zhang]{Yifan Zhang}
\email{zhangyifan95@hust.edu.cn}
\affiliation{%
  \institution{Huazhong University of Science and Technology}
  \city{Wuhan}           
  \country{China}
}

\author[Z Xue]{Ziluo Xue}
\email{xzl@hust.edu.cn}
\affiliation{%
  \institution{Huazhong University of Science and Technology}
  \city{Wuhan}           
  \country{China}
}

\author[Y Zhao]{Yanjie Zhao}
\email{yanjie_zhao@hust.edu.cn}
\affiliation{%
  \institution{Huazhong University of Science and Technology}
  \city{Wuhan}
  \country{China}
}

\author[C Fu]{Cai fu}
\email{fucai@hust.edu.cn}
\affiliation{%
  \institution{Huazhong University of Science and Technology}
  \city{Wuhan}
  \country{China}
}

\author[H Wang]{Haoyu Wang}
\email{haoyuwang@hust.edu.cn}
\affiliation{%
  \institution{Huazhong University of Science and Technology}
  \city{Wuhan}           
  \country{China}
}


\input{Sections/0.Abstract}

\maketitle

\input{Sections/1.Introduction}

\input{Sections/2.Background}

\input{Sections/3.Threatmodel}

\input{Sections/4.SMCP}
\input{Sections/5.Usecase}
\input{Sections/6.Relatedwork}

\input{Sections/7.Conclusion}


\bibliographystyle{ACM-Reference-Format}
\bibliography{acmart}

\end{document}

%% file: Sections/0.Abstract.tex
\begin{abstract}
Agentic AI systems built around large language models (LLMs) are moving away from closed, single-model frameworks and toward open ecosystems that connect a variety of agents, external tools, and resources. The Model Context Protocol (MCP) has emerged as a standard to unify tool access, allowing agents to discover, invoke, and coordinate with tools more flexibly. However, as MCP becomes more widely adopted, it also brings a new set of security and privacy challenges. These include risks such as unauthorized access, tool poisoning, prompt injection, privilege escalation, and supply chain attacks, any of which can impact different parts of the protocol workflow. While recent research has examined possible attack surfaces and suggested targeted countermeasures, there is still a lack of systematic, protocol-level security improvements for MCP. To address this, we introduce the Secure Model Context Protocol (SMCP), which builds on MCP by adding unified identity management, robust mutual authentication, ongoing security context propagation, fine-grained policy enforcement, and comprehensive audit logging. In this paper, we present the main components of SMCP, explain how it helps reduce security risks, and illustrate its application with practical examples. We hope that this work will contribute to the development of agentic systems that are not only powerful and adaptable, but also secure and dependable.
\end{abstract}

%% file: Sections/1.Introduction.tex
\section{Introduction}
\label{sec:introduction}

In recent years, rapid progress in general AI technologies, particularly advances in Large Language Models (LLMs), has shifted agentic systems away from single-model conversation paradigms. Increasingly, these systems are moving toward open ecosystems that connect heterogeneous agents, external tools, and diverse resources. LLM-powered autonomous agents now go beyond text generation but handle lots of real-world tasks by integrating modules for planning, memory, tool use, and action execution. Because of these advances, LLMs are becoming critical entities capable of continuous perception, decision-making, and action in complex environments, as demonstrated in recent studies~\cite{weng2023agent,wei2022chainofthought,yao2023treeofthoughts,shinn2023reflexion}.
At the same time, protocols designed for tool access and agent communication have become increasingly important. Mechanisms such as OpenAI function calling~\cite{functioncalling}, LangChain tool calling~\cite{LangChain}, and the Model Context Protocol (MCP) from Anthropic~\cite{mcp2024intro} provide various ways for agents to interact with a wide range of external resources. 
Among these mechanisms, the most popular and widely adopted is MCP. With this protocol, developers can use a consistent method to connect their agents to various data sources, tools, or workflows, helping keep the overall design cleaner and more modular. Instead of tying tools directly to a particular model, MCP lets various clients and models access the same set of resources. For instance, an agent can look up, choose, and use external tools simply by following a standard interface, without having to worry about the underlying details. Because of these advantages, protocols like MCP are now playing a key role in the open and interoperable agent ecosystems.

However, as MCP and similar open protocols are adopted more widely, new security and privacy concerns have emerged~\cite{hou2025mcp}. Attackers have more opportunities to target not just individual models, but the entire system, including users and external resources~\cite{GB_AIP,mcptop25}. Some of the risks that have come up include unauthorized access, poor session management, manipulation of tool metadata, and different types of injection attacks, such as command or prompt injection. These issues can show up at any stage of the protocol’s operation. For example, attackers might take advantage of user input or documents to influence what an agent does, or they could compromise an MCP server to run harmful code or gain access to sensitive information~\cite{weng2023agent,yang2025mcpsecbench,wang2025mpma,zhao2025mindyourserver,radosevich2025mcpsafety,hasan2025mcpglance,zhao2025mcpattack}.

So far, most research on MCP security has looked at where attacks might happen and what risks exist in the broader ecosystem. For example, existing work has examined issues like tool poisoning, prompt injection, and privilege escalation across servers. Some researchers have put together attack taxonomies and developed detection tools, such as MindGuard~\cite{MindGuard}, MCPLIB~\cite{MCPLIB}, and MCPTox~\cite{MCPTox}. On the agent side, the community has tried out different ways to reduce risk at the agent or tool level, which include tracking agent behavior (AgentArmor~\cite{AgentArmor}), using contract-based rules (AgentSpec~\cite{AgentSpec}), controlling privileges (SAGA~\cite{SAGA}), and isolating inputs and outputs (DRIFT~\cite{DRIFT}). Recent frameworks like A2AS~\cite{neelou2025a2asagenticairuntime} discuss runtime security controls for agentic systems, and the latest standard~\cite{GB_AIP} outlines general agent interconnection architectures.
However, there is still no dedicated protocol-level security enhancement specifically designed for MCP. Most existing solutions are either built for individual agents or applications, or they only address specific threats. Key challenges such as unified identity management, continuous security context tracking, and end-to-end policy enforcement remain largely unaddressed.

Motivated by these challenges, we introduce the Secure Model Context Protocol (SMCP). SMCP is designed as a practical security framework that makes open agent-tool environments more trustworthy, better controlled, and easier to audit from end to end. SMCP builds on the foundation of MCP, extending it with a unified digital identity and trust infrastructure, robust mutual authentication, continuous security context propagation, fine-grained policy enforcement, and comprehensive audit logging. By embedding these features directly into the protocol, SMCP addresses a wide range of security risks seen in modern agentic workflows, all while preserving interoperability and flexibility for developers.
This paper makes several key contributions:

\begin{itemize}
    \item We analyze the main security risks in agent ecosystems built around MCP, tracing threats throughout the protocol’s workflow and probing their underlying causes.
    \item We present the detailed architecture of SMCP as a security-enhanced extension to MCP, explaining how it brings together identity management, session authentication, security context propagation, policy enforcement, and audit logging into a unified protocol layer to enable consistent and robust security across agent-tool interactions.
    \item We demonstrate the effectiveness of SMCP’s built-in mechanisms in mitigating major security risks, using practical examples and systematically mapping specific threat types to the corresponding SMCP controls and enforcement strategies.
\end{itemize}

The remainder of the paper is structured as follows. \autoref{sec:backgroud} reviews the background of LLM-powered autonomous agents and the MCP. \autoref{sec:threatmodel} systematically analyzes the security risks in MCP workflows. \autoref{sec:smcp} introduces the design and mechanisms of SMCP. \autoref{sec:usecase} presents practical use cases and risk mitigation mappings. \autoref{sec:related} reviews the work related to MCP security and agentic system protection. \autoref{sec:roadmap} discusses the development roadmap and future directions. Finally, \autoref{sec:conclusion} concludes the paper.

%% file: Sections/2.Background.tex
\section{Background}
\label{sec:backgroud}

\subsection{LLM Powered Autonomous Agents}

\begin{figure}[htbp]
    \centering
    \includegraphics[width=0.78\linewidth]{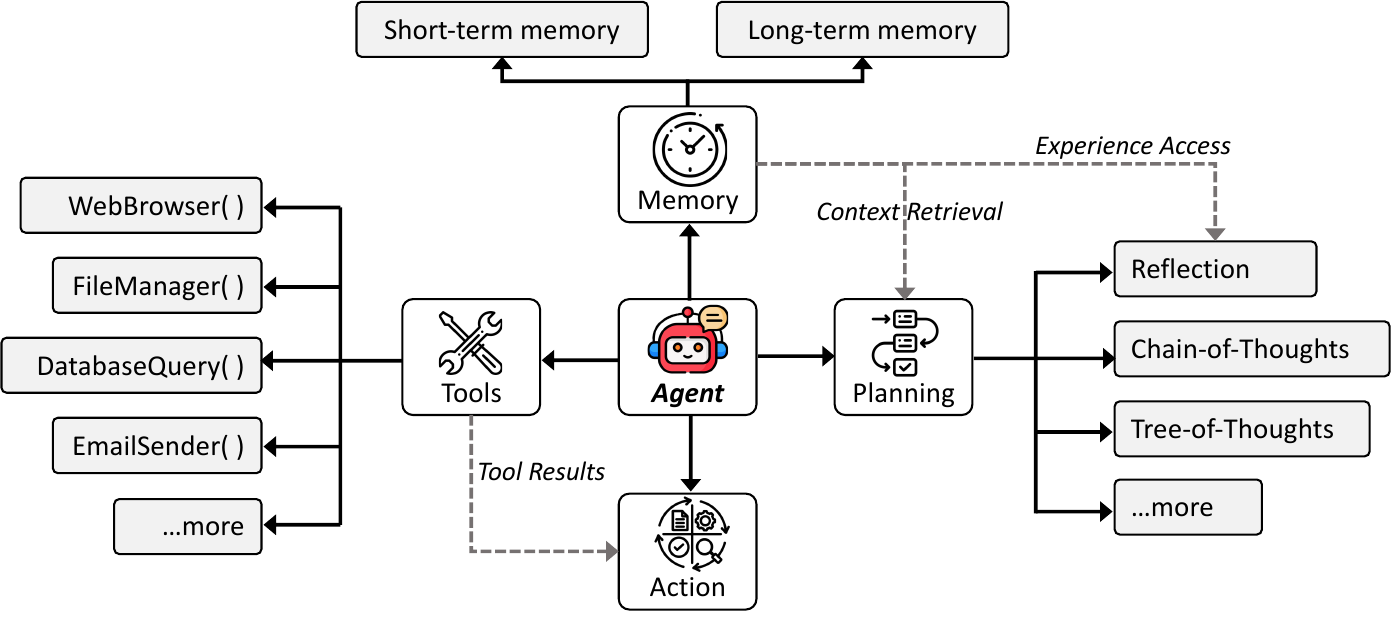}
    \caption{Overview of a Typical LLM-based Autonomous Agent Architecture.}
    \label{fig:agent}
\end{figure}

Autonomous agents powered by LLMs are an emerging direction in intelligent systems. Instead of relying solely on single-turn conversations, these agents place LLMs at the center and combine them with modules for planning, memory, and tool use (see~\autoref{fig:agent})~\cite{Gan2024Navigating,Li2024Personal}. This allows agents to understand their surroundings, make decisions, and take actions in more complex situations~\cite{weng2023agent}. As a result, LLMs can now take on long-term tasks and interact with their environment in a more active way.
Specifically, the planning module breaks down user goals into specific steps, often using strategies such as chain-of-thought reasoning or self-reflection. The memory module keeps track of both recent context and longer-term knowledge, helping the agent plan ahead and learn from experience. The tools module lets the agent reach out to external resources for tasks like searching for information, running calculations, or executing code, which greatly extends what the LLM can do. Finally, the execution module decides when to use these tools or take other actions based on the plan, and it sends any new results back to memory so the agent can keep improving over time.

\subsection{Model Context Model (MCP)}

Before the emergence of the MCP protocol, LLM agents primarily relied on a variety of non-unified approaches when interacting with external tools, as shown in \autoref{tab:tool_evolution}. While these methods were typically sufficient for small-scale, single-task scenarios, they faced significant scalability bottlenecks when required to simultaneously access a large number of heterogeneous data sources and tools or to share capabilities across multiple models and applications.

\input{Tables/tool_evolution}

As LLM agents have grown more complex and the number of available tools has increased, it has become much harder to describe, manage, and connect these resources in a way that is both scalable and independent of any single implementation. To address these issues, Anthropic introduced the Model Context Protocol (MCP) at the end of 2024~\cite{mcp2024intro}. Drawing inspiration from general-purpose protocols like the Language Server Protocol (LSP), MCP is designed as a single, flexible interface for accessing external tools and shared context.

\begin{figure}[htbp]
    \centering
    \includegraphics[width=0.95\linewidth]{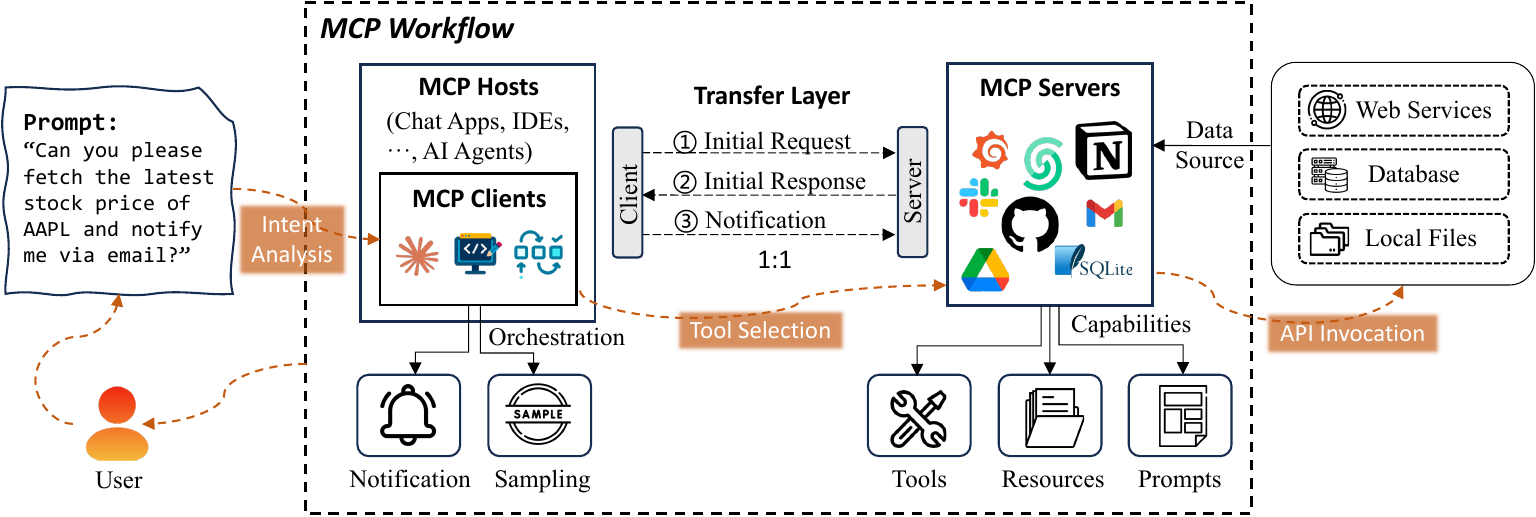}
    \caption{The Workflow of MCP (Reproduced from Our Previous Work~\cite{hou2025mcp}).}
    \label{fig:mcp}
\end{figure}

As shown in \autoref{fig:mcp}, the architecture of MCP consists of three core components: the MCP host, MCP client, and MCP server~\cite{hou2025mcp}. The MCP host is the environment where LLMs are deployed and where the task context is kept. For example, this could be a desktop assistant or an intelligent development environment. The MCP client operates within the host and serves as a bridge between the LLMs and external resources. It is responsible for starting requests, understanding user intent, managing responses, and coordinating the process of using external tools. The MCP server, which is on the service side, offers three types of capabilities for LLMs: tools, resources, and prompts. The tools module wraps external APIs or operational logic, so the client can call these tools and receive results. The resources module manages various data sources, such as databases, files, or cloud storage. The prompts module provides reusable workflow templates that help make tool use more efficient and keep interactions consistent.

The complete MCP workflow is as follows: when a user sends a request, the MCP host interprets what kind of action is required and then  the MCP clientcontacts the MCP server through a standard communication layer. The server replies with a list of available tools and capabilities that fit the current context. The host picks the right tool, makes the necessary API call, and, once the task is finished, returns the results to the user. 
This process, while ensuring generality, also retains the ability for dynamic configuration, bidirectional interaction, and multi-task scheduling, providing fundamental protocol support for building more flexible, secure, and scalable agent systems.

%% file: Tables/tool_evolution.tex
\begin{table}[htbp]
    \centering
    \caption{Comparison of LLM Tool Integration Paradigms.}
    \resizebox{1\linewidth}{!}{
    \begin{tabular}{lp{7.5cm}p{6cm}}
        \toprule[1.2pt]
        \textbf{Paradigm} & \textbf{Main Features} & \textbf{Limitations} \\
        \midrule[1.2pt]
        \multirow{2}{*}{Manual API Integration} & Developers directly connect external APIs, offering maximum flexibility for customized scenarios. & Implementation is complex, maintenance costs are high, and scalability is limited. \\
        \midrule
        \multirow{3}{*}{Standard Plugin} & Plugins are discovered and invoked through standardized interfaces such as OpenAPI, simplifying integration within one platform. & The approach is typically stateless, restricted to a single platform, and difficult to reuse across different environments. \\
        \midrule
        \multirow{3}{*}{Agent Framework} & Tools are abstracted and orchestrated within an agent framework (e.g., LangChain), supporting dynamic selection and workflow management by the LLM. & Integration is highly dependent on specific frameworks, resulting in weak interoperability and limited extensibility. \\
        \midrule
        \multirow{3}{*}{RAG + Vector DB} & Retrieval-augmented generation enables LLMs to incorporate external knowledge via vector search, enhancing context understanding. & This method is limited to information retrieval and does not support action execution or data modification. \\
        \midrule
        \multirow{3}{*}{Protocol-based (MCP)} & A unified and extensible protocol enables decoupled, dynamic discovery and invocation of tools and resources across applications and models. & The protocol ecosystem is still evolving, and consistency and security best practices are not yet fully established. \\
        \bottomrule[1.2pt]
    \end{tabular}}
    \label{tab:tool_evolution}
\end{table}

%% file: Sections/3.Threatmodel.tex
\section{Security Risks of MCP}
\label{sec:threatmodel}

\begin{figure}[htbp]
    \centering
    \includegraphics[width=1\linewidth]{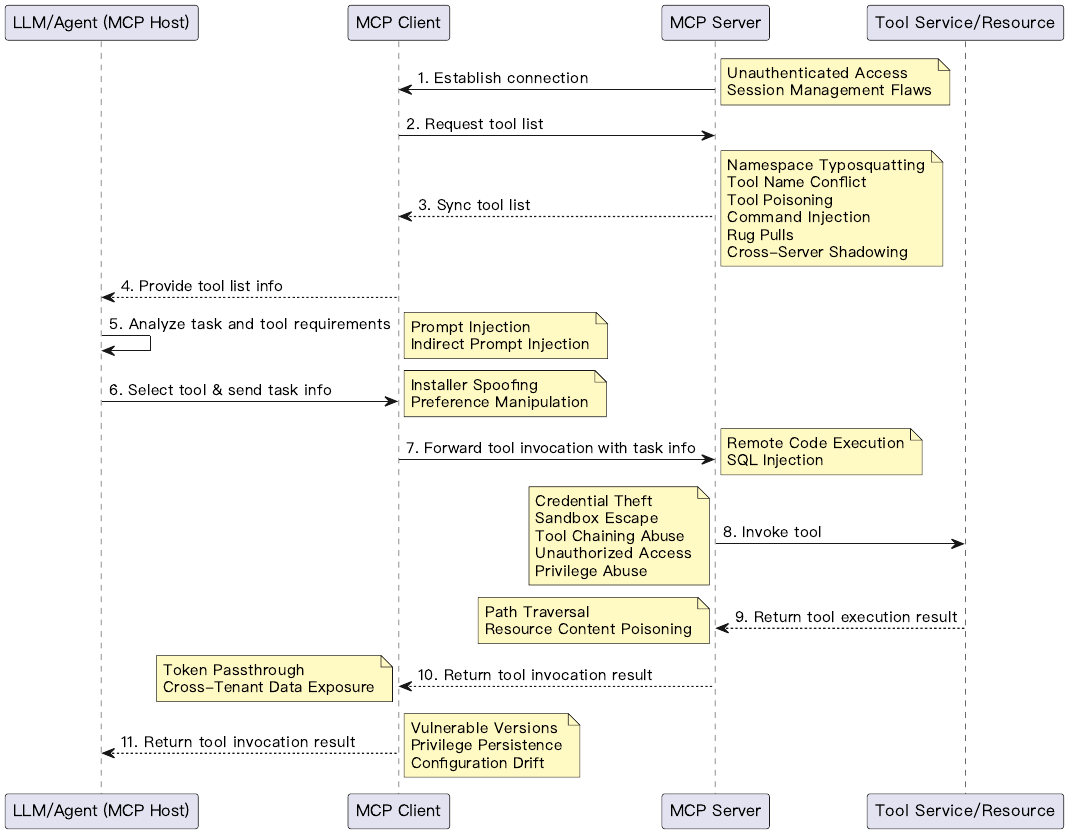}
    \caption{Security Threats and Attack Surfaces in MCP Workflow.}
    \label{fig:risk}
\end{figure}

The openness and flexibility that MCP offers for integrating AI agents and external tools also introduce a wide range of security and privacy risks. As illustrated in \autoref{fig:risk}, threats may appear throughout the entire MCP workflow, and each stage can present its own set of vulnerabilities that attackers might exploit~\cite{yang2025mcpsecbench,zhao2025mindyourserver,hasan2025mcpglance,mcptop25}. These risks may come from various sources, including outside attackers, malicious or compromised developers, supply chain participants, or insiders with elevated access~\cite{hou2025mcp}. Attackers are usually interested in gaining unauthorized access, stealing data, influencing agent behavior, or compromising the protocol’s integrity. Each part of the MCP ecosystem, such as the host, client, server, and connected tools, may become a target for attacks.

In our analysis, we examine the end-to-end workflow and consider possible risks at each stage. Early steps in the process may be affected by weak authentication or poor session management, while later phases may be exposed to tool poisoning, prompt injection, fake installers, or remote code execution. Other threats, such as stolen credentials, privilege escalation, leaked tokens, and misconfigurations, can also occur as the workflow progresses. In many cases, attackers may combine several risks to increase the overall impact.
\autoref{tab:security_risk} summarizes the main types of risk and how they can appear in different stages of the workflow. Since our work is not focused on providing an exhaustive list of all possible threats, we do not discuss each type in detail. Instead, our main concern is the protocol’s design and the enforcement mechanisms that can help reduce these risks.

\input{Tables/security_risk}


%% file: Tables/security_risk.tex
\begin{table}[htbp]
    \centering
    \caption{Summary of Security Risks Across MCP Workflow Steps (Refined from our previous work~\cite{hou2025mcp}).}
    \resizebox{1\linewidth}{!}{
    \begin{tabular}{cllp{8.8cm}}
        \toprule[1.2pt]
        \textbf{Step} & \textbf{Risk Name} & \textbf{Risk Source} & \textbf{Risk Description} \\
        \midrule[1.2pt]
        \multirow{4}{*}{1} 
            & Unauthenticated Access         & Protocol/Design Flaw          & Lack of robust authentication allows attackers to impersonate clients or servers, leading to unauthorized access and potential data leakage. \\  
            & Session Management Flaws       & Protocol/Design Flaw       & Weak session handling enables hijacking, replay attacks, or persistent unauthorized access by failing to securely generate, expire, or revoke sessions. \\
        \midrule
        \multirow{12}{*}{3} 
            & Namespace Typosquatting        & Malicious Developer & Attackers register servers with deceptively similar names to legitimate ones, tricking users or agents and facilitating supply chain compromise. \\
            & Tool Name Conflict             & Malicious Developer & Ambiguous or conflicting tool names enable attackers to substitute malicious tools for legitimate ones during selection. \\
            & Tool Poisoning                 & Malicious Developer & Malicious logic or hidden instructions are embedded in tool metadata, causing agents to execute unintended or harmful operations. \\
            & Command Injection              & External Attacker          & Insecure parameter handling allows attackers to inject system commands, leading to arbitrary code execution on the server. \\
            & Rug Pulls                      & Malicious Developer & Initially benign servers or tools are later updated to include malicious payloads, betraying established trust. \\
            & Cross-Server Shadowing         & Malicious Developer & Attackers exploit overlapping tool definitions to cause agents to invoke attacker-controlled tools instead of trusted ones. \\
        \midrule
        \multirow{4}{*}{5} 
            & Prompt Injection               & User/Context Injection     & Malicious user prompts manipulate the model’s reasoning, triggering unintended actions or privilege escalation. \\
            & Indirect Prompt Injection      & External Attacker     & Adversarial instructions are embedded in external data sources, covertly influencing model behavior or leaking data. \\
        \midrule
        \multirow{4}{*}{6} 
            & Installer Spoofing             & External Attacker & Attackers distribute tampered installer packages or auto-installers, introducing malware or backdoors during tool setup. \\
            & Preference Manipulation        & Malicious Developer     & Persuasive or deceptive tool descriptions bias agent or model selection, steering preference toward attacker-controlled tools. \\
        \midrule
        \multirow{4}{*}{7} 
            & Remote Code Execution          & External Attacker          & Insufficient input validation allows attacker-supplied data to trigger arbitrary code execution on the server. \\
            & SQL Injection                  & External Attacker          & Unsanitized parameters are passed into database queries, enabling manipulation or exfiltration of backend data. \\
        \midrule
        \multirow{10}{*}{8} 
            & Credential Theft               & Insider/Privilege Abuse    & Attackers extract API keys, tokens, or credentials from the execution environment for long-term unauthorized access. \\
            & Sandbox Escape                 & User/Context Injection & Malicious tools escape isolation, gaining access to host system resources or enabling lateral movement. \\
            & Tool Chaining Abuse            & User/Context Injection & Multiple low-risk tools are chained to perform unintended high-impact operations, evading policy checks. \\
            & Unauthorized Access            & Insider/Privilege Abuse    & Insufficient access control allows invocation of tools or resources beyond intended permissions. \\
            & Privilege Abuse                & Insider/Privilege Abuse    & Over-permissioned tools or misconfigured policies enable persistent unauthorized operations or escalation. \\
        \midrule
        \multirow{4}{*}{9} 
            & Path Traversal                 & External Attacker          & Improper path handling allows attackers to access or overwrite arbitrary files via crafted return data. \\
            & Resource Content Poisoning     & External Attacker          & Manipulated output data injects malicious content, misleading downstream agents or corrupting workflows. \\
        \midrule
        \multirow{4}{*}{10} 
            & Token Passthrough              & Protocol/Design Flaw       & Sensitive tokens or credentials are inadvertently forwarded along the workflow, exposing them to unauthorized parties. \\
            & Cross-Tenant Data Exposure     & Insider/Privilege Abuse    & In multi-tenant environments, improper segregation leads to leakage of data between tenants or sessions. \\
        \midrule
        \multirow{5}{*}{11} 
            & Vulnerable Versions            & Version/Config Management  & Outdated or unpatched tool/server versions remain in use, exposing the system to known exploits. \\
            & Privilege Persistence          & Version/Config Management  & Revoked or outdated credentials are not invalidated after updates, allowing retention of unauthorized access. \\
            & Configuration Drift            & Version/Config Management  & Manual or uncoordinated configuration changes accumulate, causing deviation from the intended security baseline. \\
        \bottomrule[1.2pt]
    \end{tabular}}
    \label{tab:security_risk}
\end{table}

%% file: Sections/4.SMCP.tex
\section{Secure Model Context Protocol (SMCP)}
\label{sec:smcp}

As stated before, the security threats in open agent-tool ecosystems are deeply rooted in the lack of unified trust, fragmented authentication, and limited auditability throughout the workflow. 
Addressing such systemic challenges requires a holistic, protocol-level solution that can \textbf{guarantee the authenticity of all participants, enforce strong mutual authentication, maintain a continuous security context across all operations, and ensure comprehensive accountability}.

\begin{figure}[htbp]
    \centering
    \includegraphics[width=0.95\linewidth]{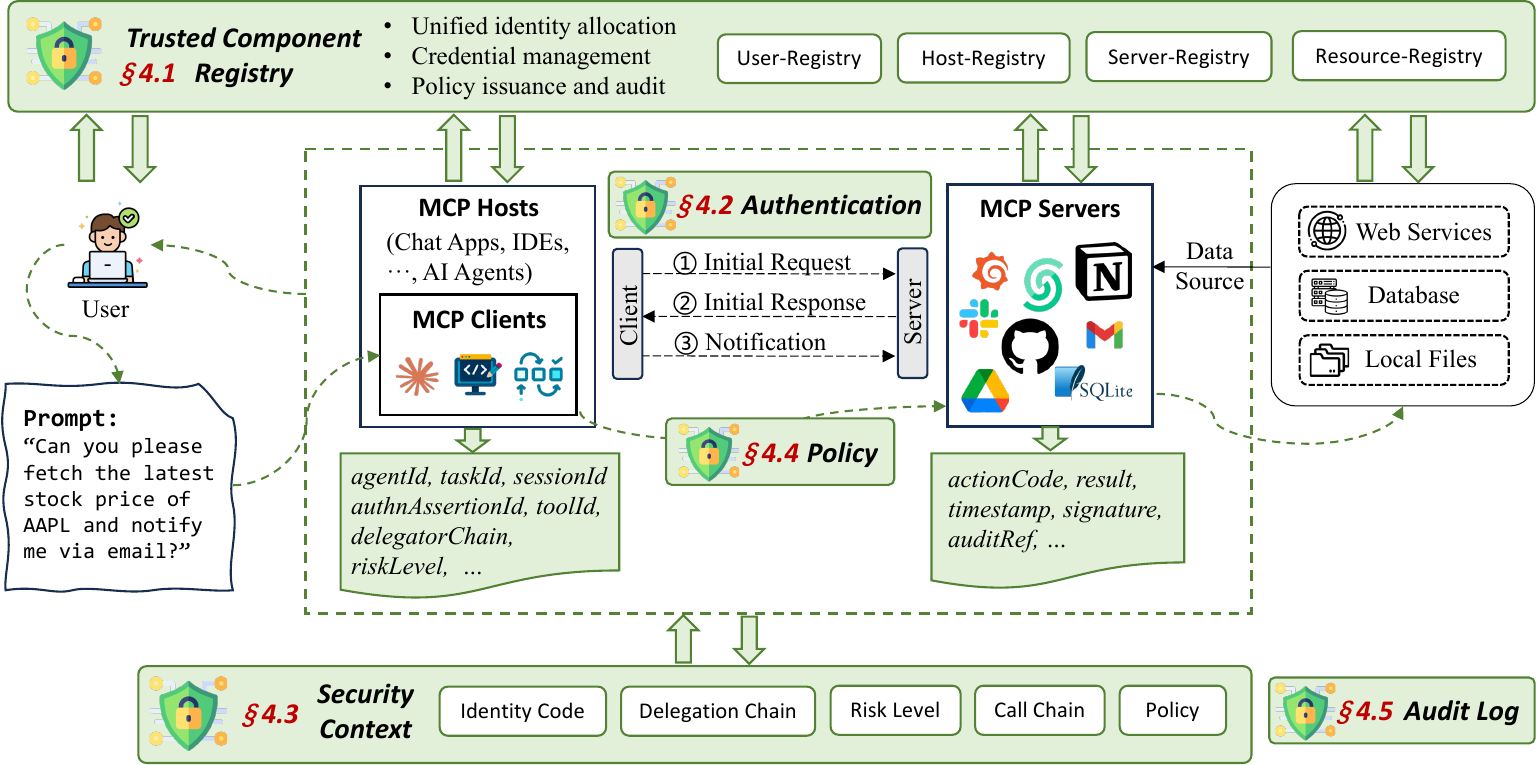}
    \caption{SMCP: Secure Model Context Protocol.}
    \label{fig:smcp}
\end{figure}

SMCP is designed to meet these needs. It serves as a unified security framework that combines identity management, session authentication, security context tracking, policy enforcement, and audit logging. At the core of this approach, the \emph{Trusted Component Registry} makes sure that only verified and traceable participants are allowed in agent-tool interactions. Each connection is created using mutual authentication, and every operation is linked to a security context that contains details about identity, delegation, and risk. The system enforces detailed access policies to support minimal privilege and flexible protection. Complete audit logs are kept so that all important actions can be followed from start to finish.
As shown in~\autoref{fig:smcp}, SMCP brings these key elements together to provide strong access control, protection during operation, and clear accountability. This allows multiple agents to work together securely and reliably, while still keeping the system compatible and flexible.

\subsection{Trusted Component Registry}
\label{subsec:registry}

\subsubsection{Unified Digital Identity and Trust Infrastructure}

SMCP is built on a unified digital identity and trust infrastructure that spans multiple systems and trust domains, which forms the foundation for protocol operation. Unlike solutions that only address a single type of entity, such as agents, SMCP brings together all key participants in the ecosystem. This includes not only human users and organizational accounts, but also different types of agents, tools, and external resource services such as MCP servers, model services, and dataset services. It also covers governance components like identity and policy services, bringing them all under a common identity and trust framework.  This design covers not only the structured encoding and account management of all entity types, but also a comprehensive \emph{Trusted Component Registry} and standardized assertion mechanisms: only those components registered and bound to the unified identity namespace via digital credentials are recognized as trusted components eligible to participate in the SMCP invocation plane.

\subsubsection{Scope and Key Elements}

The trust infrastructure of SMCP comprises three interrelated parts: 
\begin{itemize}
    \item Structured identity codes for all principal types;
    \item Identity accounts and digital credential managed by the \emph{Trusted Component Registry};
    \item Standardized authentication assertions and pre-access verification processes.
\end{itemize}
These components provide a consistent trust anchor throughout the operation of SMCP, supporting session establishment, capability access, and call-chain auditing. \autoref{tab:entity} summarizes the major entity types and key attributes managed within this infrastructure.

\input{Tables/entity}

\subsubsection{Structured Identity Code for All Entities}

\begin{figure}[htbp]
    \centering
    \includegraphics[width=0.7\linewidth]{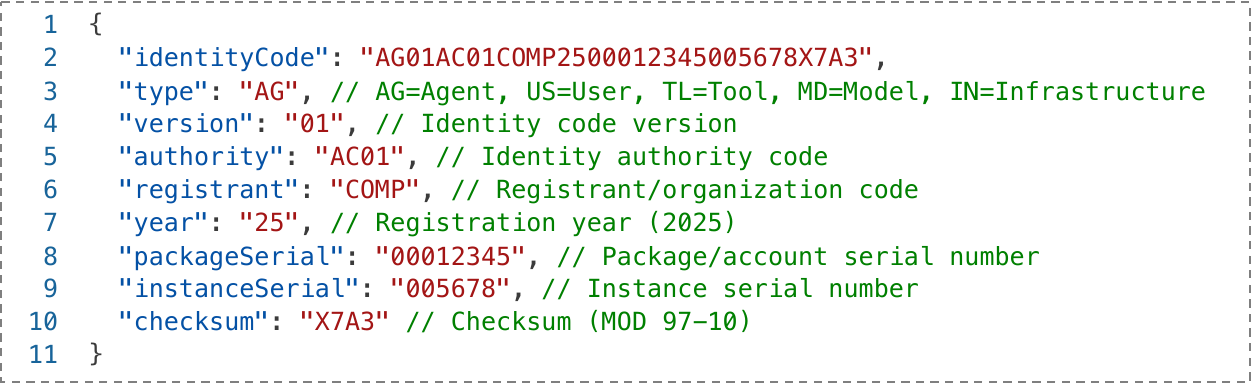}
    \caption{Example of SMCP Structured Identity Code.}
    \label{fig:identify}
\end{figure}

At the identification layer, SMCP defines unified encoding rules and type fields for different entity categories.An example as shown in \autoref{fig:identify}. Every discoverable, callable, or delegable entity is assigned a fixed-length, 32-character identity code using the character set [0-9A-Z]. This applies to users, agents, tool services, models, and resource endpoints. The code reserves type fields for distinguishing users (human or organizational), agents (general or domain-specific), tools and resources (MCP servers, APIs, model/data services), and governance/audit components (identity services, policy points, log aggregators, etc). Other fields encode version, authority, registrant, registration year, package/account serial, instance serial, and a checksum. This cross-type, parseable encoding enables global uniqueness, rapid validation (e.g., MOD 97-10 checksum), and binding of all protocol interactions to a stable identity namespace without exposing internal implementation details.

\subsubsection{Identity Lifecycle and Credential Management}

SMCP defines a unified identity lifecycle management framework atop the \emph{Trusted Component Registry}, including registration, verification, archival, credential issuance, and activation for each entity type. For agents, the sponsor or developer submits a registration request with functional descriptions, system composition and environment digests (e.g., code/model hashes, dependency fingerprints), behavioral boundaries, and supporting documentation. The registration service evaluates risk and required evidence based on task sensitivity and deployment context, verifies all proofs, then creates a registry entry, assigns the identity code, and snapshots verified information as the baseline.

Likewise, users and organizations must complete registration and verification (potentially leveraging existing IAM/IDaaS infrastructure), incorporating account attributes, organizational relationships, and baseline authorization into the unified identity domain. Tool and resource services must register before being added to the capability directory or invoked by SMCP, disclosing provider/operator, capability sensitivity, and compliance boundaries. Models and datasets must declare training data sources, usage restrictions, and sensitivity tags. This creates a unified graph of users, agents, tools, models, data, and organizations, enabling delegation chains, access control, and audit trails.

Once registered, independent credential issuance services generate digital credentials for each entity in the form of public key certificates or verifiable credentials containing the identity code, type, provider/delegator information, capability scope, privilege level, security/compliance attributes, and public key reference. After issuance, the identity code and credential are returned for secure storage and use. Only after this process is an entity's digital identity considered active and eligible for trusted participation in SMCP operations.
To prevent ``register-once, never-update'' risks, SMCP supports unified update, lock, and revocation mechanisms. When an agent’s implementation, capability, or environment changes significantly, an identity update process is triggered (identity code remains, registry/account info is updated, credentials may be rotated). Tool/resource services follow similar flows for capability/sensitivity/operator changes. For misbehavior, credential leaks, or delegator revocation, identity accounts may be locked or deactivated, with full audit trails preserved for compliance and forensic review.

\subsection{Session Establishment and Mutual Authentication}
\label{subsec:authentication}

A secure session in SMCP is established through a process of mutual authentication between entities, based on the unified digital identity and credential infrastructure provided by the \emph{Trusted Component Registry}. This approach ensures that both parties in a session can reliably verify each other's identities. Each session is then linked to a verifiable security context, which serves as the basis for all secure interactions and the transfer of context information throughout the protocol.

\subsubsection{Authentication and Mutual Verification}

During cross-entity interactions, trust is established not by request payloads alone, but via standardized authentication and assertion flows:

\begin{itemize}
  \item When a subject initiates a request, the relying party evaluates its access control policy to determine the required authentication strength (e.g., agent identity only, or additional tool/model trust attributes).
  \item Upon challenge, the requester assembles an appropriate credential bundle, signs dynamic challenges, and presents them (e.g., via mutual TLS or application-level assertions).
  \item The verifier (identity service or relying party) checks credential signatures, trust chains, validity, revocation, and ensures scope matches the request. For delegation, registry records are used to validate the full authorization chain from delegator to current actor.
  \item After verification, an authentication assertion is generated, including result, assurance level, subject identity, delegator summary, unique ID, and timestamp. The relying party uses this assertion and local policy to decide on access, privileges, further checks, or session setup.
\end{itemize}

With unified digital identity, a \emph{Trusted Component Registry}, and trust infrastructure across all entity types, SMCP ensures that all protocol-level identifiers (\texttt{agentId}, \texttt{toolServiceId}, \texttt{modelId}, \texttt{datasetId}, etc.) can be traced to a registered, auditable trusted component. This provides a consistent, measurable trust layer for all cross-boundary interactions.

\subsubsection{Session Initialization and Security Context Binding}

After mutual authentication is successfully completed, a dedicated session is established between the interacting entities. At this stage, SMCP generates a session-specific security context that aggregates essential security semantics, including the identities of both parties, the verified delegation chain, risk level, session identifiers, and relevant metadata. This security context receives a unique identifier, such as \texttt{sessionId} or \texttt{callChainId}, and is cryptographically bound to the session using the secure key material negotiated during authentication.
The security context is explicitly associated with all subsequent messages, task requests, and tool invocations within the session. This association ensures that every operation can be traced back to the original authentication event and that context integrity is maintained throughout the invocation chain. If the session is interrupted, expires, or is terminated, the associated security context is invalidated to prevent unauthorized reuse. Through this mechanism, all downstream operations inherit a consistent and auditable security baseline, supporting end-to-end access control, policy enforcement, and auditability across the entire SMCP workflow.


\subsection{Security Context}
\label{subsec:context}

\subsubsection{Definition and Design Principles}

Building on digital identity and trust infrastructure, SMCP defines a unified security context structure at the protocol abstraction layer and requires that this context be explicitly carried by or be traceable from core interaction elements such as sessions, tasks, messages, and tool invocations. The unified security context consolidates security semantics that are scattered across different components and boundaries into a set of common fields, thus providing a shared language at the protocol level for end-to-end authentication, authorization control, and audit tracing. \autoref{tab:context} presents the main fields of this security context.

\input{Tables/context}

This unified security context is not intended to replace the field definitions in existing standards. Instead, it integrates with current structures in two ways. First, for existing session, task, and message structures such as \texttt{sessionId}, \texttt{taskId}, and \texttt{id} in agent interaction standards, SMCP attaches references or digests of the security context by extending metadata or security extension fields. Second, for tool capability invocations and responses, SMCP separates the security context into invocation metadata, which is transmitted together with the tool invocation payload. Through this mechanism, whether for model-oriented context construction, resource-oriented capability access, or long-chain invocations involving multiple components, all events can be represented as occurring under a unified security context and endorsed by the \emph{Trusted Component Registry}. This approach provides a consistent security semantic foundation for subsequent access control, risk management, and audit tracing.

\subsubsection{Roles and Objects in SMCP}
The runtime environment of SMCP involves four core roles:

\begin{itemize}
\item \textbf{Invoking Agent}: This includes both entry-point agents and downstream business agents. These agents initiate capability invocations via the tool access component, with their identities mapped to corresponding \texttt{agentId} and trust metadata within the SMCP identity system.
\item \textbf{Tool Service}: This component connects to the tool catalog and runtime environment, is responsible for receiving invocation requests, orchestrating specific tool instances, and aggregating results. It is typically identified externally by a \texttt{toolServiceId}.
\item \textbf{Tool Instance}: These are the operational units that perform actual actions within the resource access domain, such as database queries, API calls, or local script executions.
\item \textbf{Policy Engine}: This component works in conjunction with identity and authorization services. It makes decisions on authorization, downgrade, or denial based on the security context, tool security properties, and relevant component metadata.
\end{itemize}

Building on these roles, the protocol revolves around three main types of objects: \textbf{Tool Capability Description}, \textbf{Tool Invocation Request}, and \textbf{Tool Execution Response}, as illustrated in \autoref{tab:object}. SMCP requires explicit embedding of security context fields (such as \texttt{callChainId}, \texttt{delegatorChain}, \texttt{riskLevel}, etc.) within these objects, and associates them with the identity and capability metadata of each component. This ensures that the complete invocation chain, from user to agent to tool, maintains a consistent security semantic throughout. 

\input{Tables/object}

Through these objects, SMCP brings the high-level security semantics down to the tool layer, so that each invocation can be verified and audited in terms of who authorized it, the associated risk level, and what capability and underlying components have been accessed.

\subsubsection{Security Extensions for Tool Capabilities}
\begin{figure}[htbp]
    \centering
    \includegraphics[width=1\linewidth]{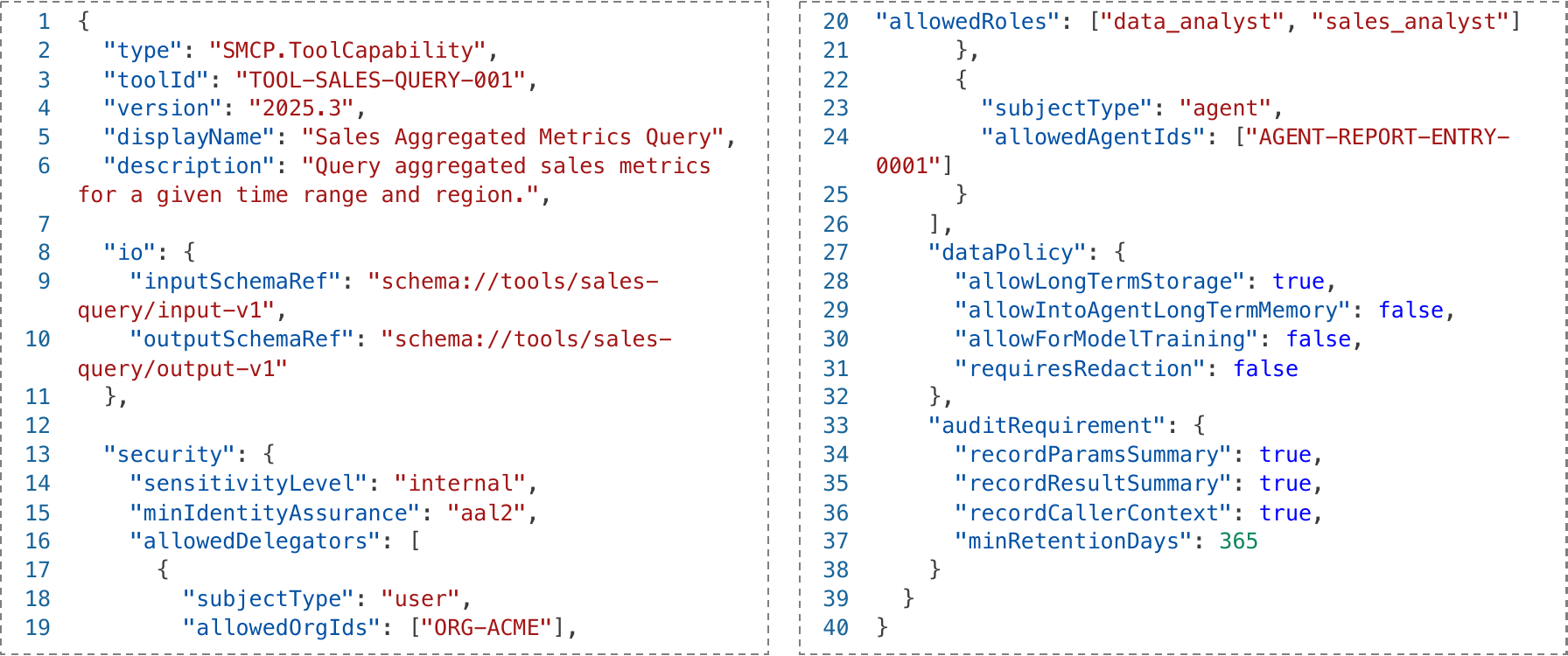}
    \caption{SMCP Tool Capability.}
    \label{fig:tool_cap}
\end{figure}

Building upon existing tool capability descriptions (such as name, input/output patterns, invocation quotas, etc.), SMCP introduces a set of security attributes for each tool to drive authorization decisions and invocation downgrade strategies. \autoref{tab:capacity} presents an illustrative set of such fields. 
\autoref{fig:tool_cap} is an example of a tool capability description with the security extension fields included alongside the standard capability metadata:

\input{Tables/capacity}

Tool services can fetch and cache such security capability descriptions when loading tool catalogs. Subsequently, each time an SMCP tool invocation request is received, the tool service matches the security context in the invocation message with these fields, thereby achieving access control and risk evaluation in a unified semantic framework.

\subsubsection{Security Semantics of SMCP Invocation and Response}

SMCP treats a tool invocation as a controlled capability access request, issued by an agent on behalf of an upstream subject, within a specific delegation chain and risk configuration. ~\autoref{tab:field} summarizes the key security fields in SMCP invocation and response messages.

\input{Tables/field}

During the invocation phase, the agent constructs an SMCP tool invocation request based on the security context formed in the current session, binding task semantics with tool capabilities, security constraints, and component identities. 
Upon completing the execution of a specific tool instance, the tool service returns the result and relevant security metadata to the invoking agent through an SMCP tool execution response message.

\subsection{Policy Enforcement and Least-Privilege Control} 
\label{subsec:policy}

SMCP leverages the unified security context and capability model to support fine-grained policy enforcement and the principle of least privilege throughout the tool invocation lifecycle. Every invocation request is evaluated by the policy engine, which makes decisions based on attributes such as agent identity, delegator chain, risk level, data sensitivity, and permitted operations.

\subsubsection{End-to-End Policy Enforcement Workflow}
As shown in~\autoref{fig:sequence}, SMCP policy enforcement is tightly integrated into the end-to-end tool invocation workflow:

\begin{figure}[htbp]
    \centering
    \includegraphics[width=1\linewidth]{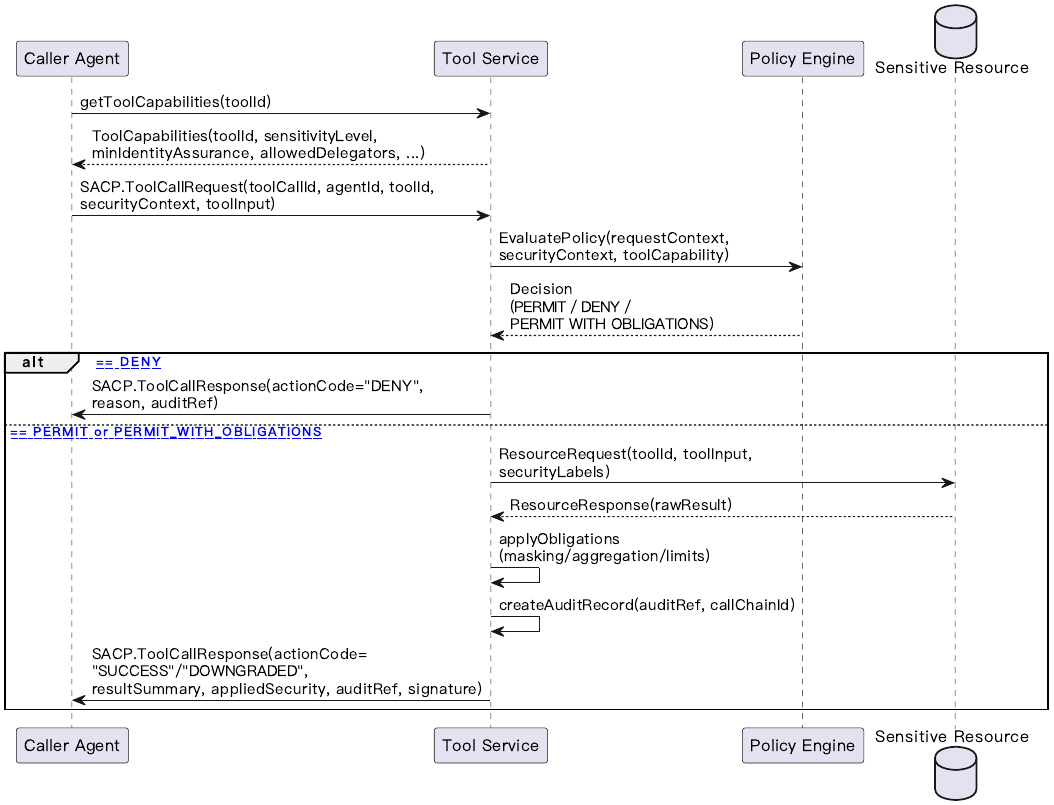}
    \caption{Sequence Diagram of Secure Agent~$\leftrightarrow$~Tool Interoperability in SMCP.}
    \label{fig:sequence}
\end{figure}

\begin{enumerate}
    \item \textbf{Capability Discovery:} The caller agent initiates by retrieving tool capability descriptions, which include both functional and security-relevant metadata (such as sensitivity level, allowed delegators, and identity assurance thresholds).
    \item \textbf{Invocation Request Construction:} The agent constructs a tool invocation request, embedding the comprehensive security context (e.g., \texttt{callChainId}, \texttt{delegatorChain}, \texttt{riskLevel}, \texttt{dataSensitivity}, and identity fields).
    \item \textbf{Policy Evaluation:} Upon receiving the request, the tool service forwards the relevant security context and tool capability information to the policy engine.
    \item \textbf{Decision Making:} The policy engine evaluates the request against configured authorization policies, considering both the runtime security context and the static properties of the tool. It returns a decision of \texttt{PERMIT}, \texttt{DENY}, or \texttt{PERMIT WITH OBLIGATIONS} (such as masking, aggregation, or rate limits).
    \item \textbf{Enforcement and Obligation Application:} If permitted, the tool service executes the tool call, applying any obligations imposed by the policy engine. If denied, a detailed denial response is returned, including an audit reference.
    \item \textbf{Audit and Response:} All actions, including policy decisions and applied obligations, are logged with references to the security context for subsequent auditing. The agent receives the response, which includes both result data and the relevant security and audit metadata.
\end{enumerate}

This sequence ensures that every tool invocation is subject to dynamic, context-aware policy evaluation, and that the minimum required permissions are strictly enforced at each step. The explicit embedding of audit references and security outcomes in the response messages further supports traceability and compliance.

\subsubsection{Policy Evaluation Mechanisms and Enforcement Architecture}
The policy enforcement architecture of SMCP is fundamentally anchored in the propagation of an explicit, structured security context with every invocation across the protocol chain. Within this framework, the policy engine acts as the Policy Decision Point (PDP), while agents and tool services function as Policy Enforcement Points (PEPs). This separation of roles enables a modular and auditable security model, ensuring that policy enforcement remains consistent and traceable across diverse deployment environments.
Unlike traditional static access control, SMCP evaluates each tool invocation dynamically based on the complete security context. This context includes the full delegation chain, risk level, data sensitivity, and the identity and trust level of every participant in the call chain. The policy language enables fine-grained constraints on any of these attributes; for instance, a policy can permit access to a sensitive function only if the risk level is below “medium” and the delegation chain contains a verified human. Beyond simple allow or deny outcomes, policies may impose obligations such as result masking, rate limiting, or requiring additional verification steps, all tightly bound to context attributes. Every decision and enforcement action is linked to unique identifiers in the context, supporting end-to-end traceability and compliance.
\begin{figure}[htbp]
    \centering
    \includegraphics[width=\linewidth]{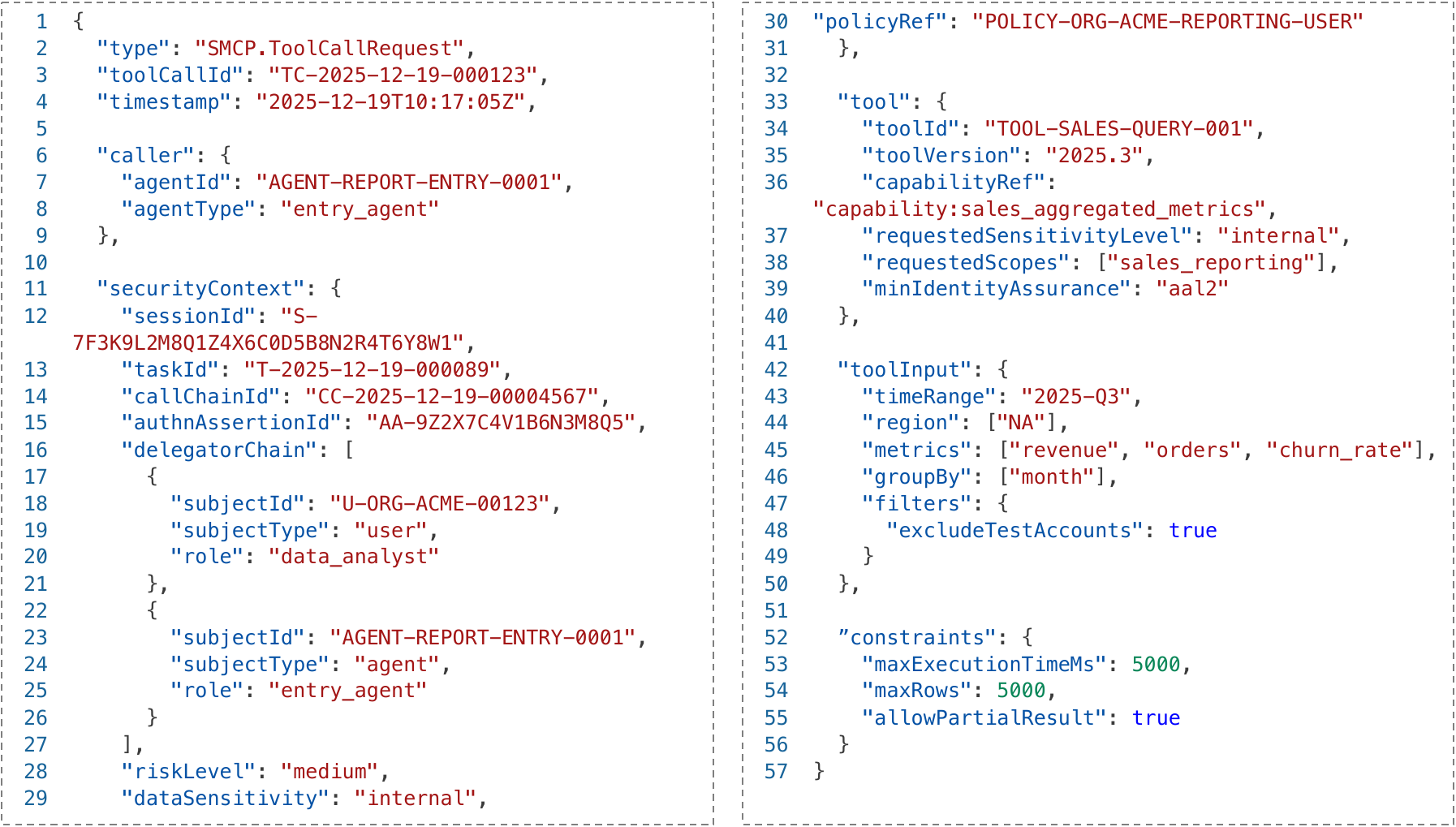}
    \caption{Example of SMCP Tool Invocation Request. This includes delegation chain, risk level, data sensitivity, and capability constraints, which are evaluated by the policy engine to determine access and obligations.}
    \label{fig:request}
\end{figure}

\autoref{fig:request} illustrates a typical SMCP tool invocation request. Key fields such as \texttt{delegatorChain}, \texttt{riskLevel}, and \texttt{dataSensitivity} provide the basis for policy decisions by the engine, which determines whether the action is permitted, downgraded, or denied. The \texttt{agentId} and \texttt{toolId} fields associate the invocation with specific agent and tool entities, supporting both access control and audit requirements. In this example, the policy engine can enforce that only requests with an appropriate risk level and an allowed delegation chain are processed, and may require masking or other obligations if the data sensitivity demands it.

\begin{figure}[htbp]
    \centering
    \includegraphics[width=\linewidth]{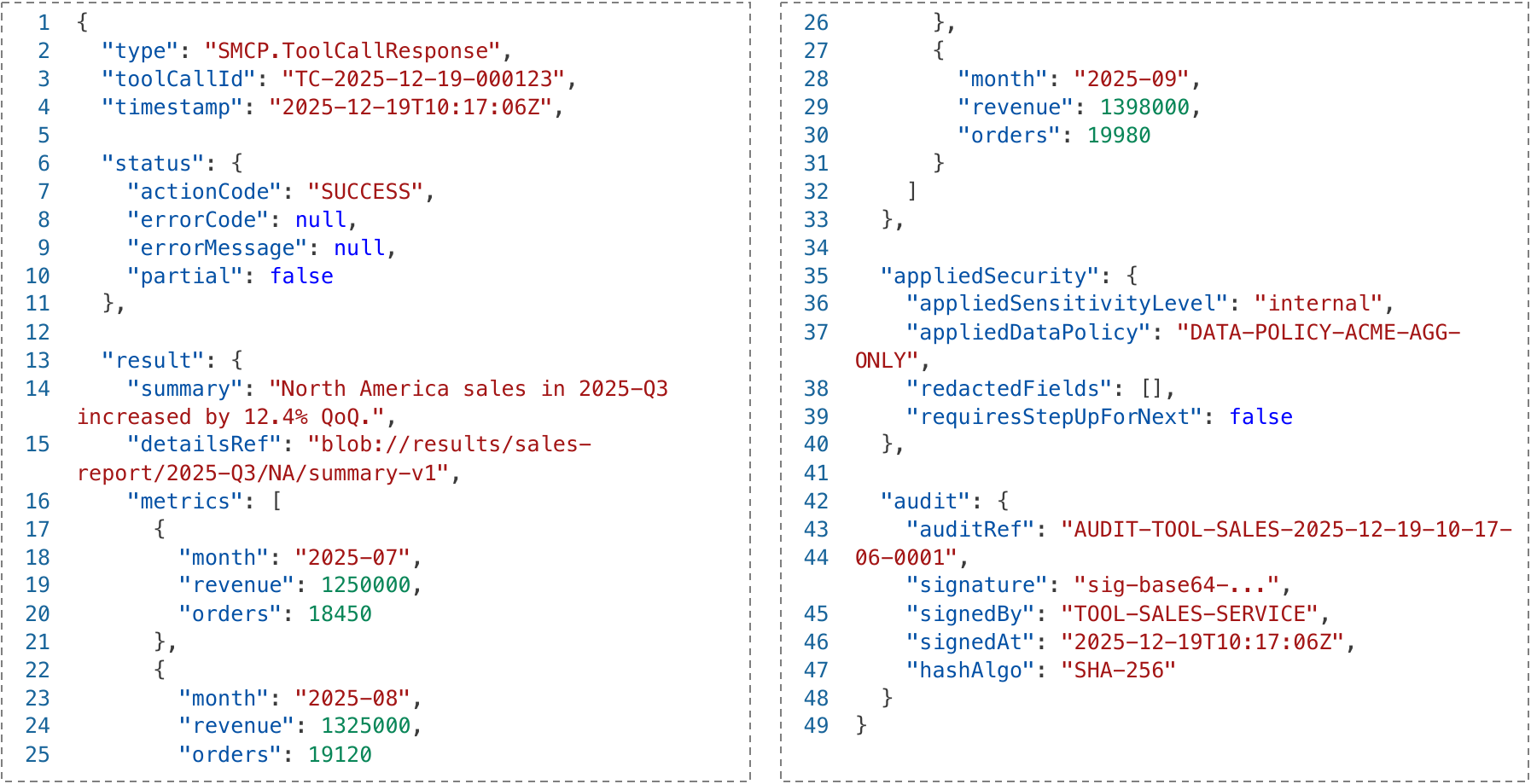}
    \caption{Example of SMCP Tool Execution Response. The response details the applied security policy, actual data handling, and provides audit and signature fields to enable end-to-end traceability and compliance.}
    \label{fig:response}
\end{figure}

Once the policy engine reaches a decision, the tool service must enforce the result, including any specified obligations. As shown in \autoref{fig:response}, the tool execution response records the applied security policy, data handling status, and audit information. The \texttt{appliedSecurity} field specifies the sensitivity level and data policy that were enforced, while \texttt{auditRef} and \texttt{signature} ensure the result’s authenticity and end-to-end traceability. This structured response allows downstream systems and auditors to verify compliance with organizational and regulatory policies.

Through this context-driven, fine-grained policy evaluation, SMCP enforces the principle of least privilege: at every invocation, only the minimal necessary permissions and capabilities are granted, and these are dynamically adjusted as context changes, ensuring robust and adaptive security.

\subsection{Audit Logging and Traceability}
\label{subsec:log}

Comprehensive audit logging is a cornerstone of SMCP's security and compliance framework. Every critical protocol operation, including agent authentication, capability negotiation, policy evaluation, tool invocation, and response delivery, is systematically captured in the audit log with structured and high-fidelity metadata. Rather than serving as a passive record, the audit log is designed to actively support retrospective security analysis, such as detecting misconfigurations or abuse, as well as demonstrating compliance with regulatory or organizational policies.
A distinctive feature of SMCP audit logging is its close integration with the security context that is propagated throughout each protocol interaction. Each log entry is bound to unique identifiers such as \texttt{callChainId} and \texttt{auditRef}, precise timestamps, digital signatures, and a complete contextual snapshot of the operation. This approach ensures comprehensive, end-to-end traceability and guarantees non-repudiation for every action within the system.

As summarized in \autoref{tab:log}, the audit log captures a broad range of key attributes, including the identities of all actors and subjects, detailed operation metadata, policy decisions and obligations, sensitivity and risk levels, result summaries, and cryptographic proofs of integrity. This comprehensive coverage enables the audit log to support real-time monitoring, automated risk detection, and post-incident forensic analysis, forming the foundation for both proactive and retrospective security controls.
\input{Tables/log}

\begin{figure}[htbp]
    \centering
    \includegraphics[width=1\linewidth]{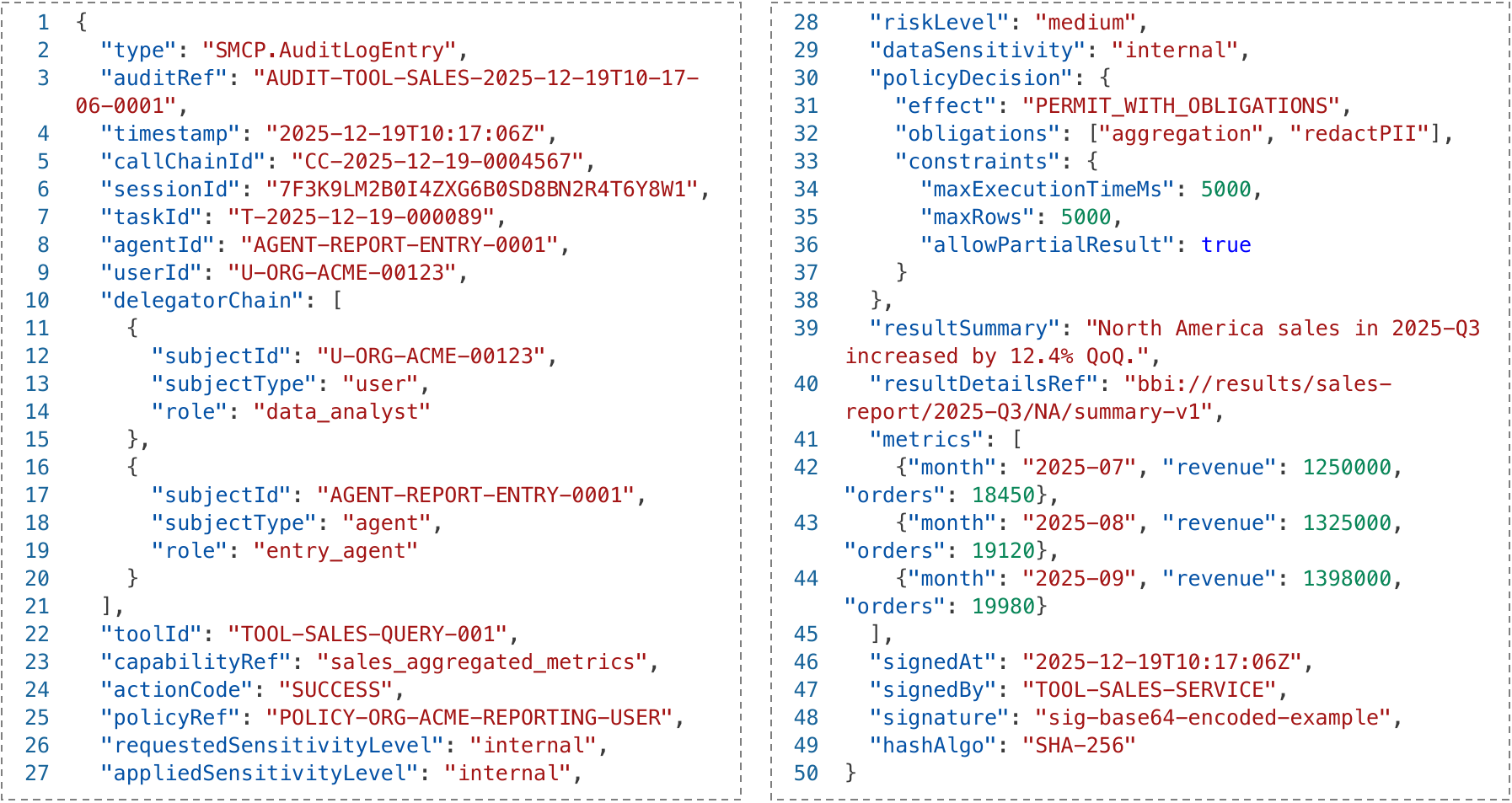}
    \caption{Example of a Structured Audit Log Entry in SMCP.}
    \label{fig:log}
\end{figure}

An example of a structured audit log entry is shown in \autoref{fig:log}, which demonstrates how these attributes are organized in practice. Each entry can be cross-referenced with both the originating request and the resulting response, ensuring a complete and verifiable chain of custody for sensitive operations. By mandating explicit, structured, and tamper-evident audit records for all security-relevant events, SMCP provides trustworthy and contextually justified decision-making in both automated and human-in-the-loop agentic workflows.

%% file: Tables/entity.tex
\begin{table}[htbp]
  \centering
  \caption{Entity Types and Key Attributes in the SMCP Trusted Component Registry}
  \label{tab:entity}
  \resizebox{1\linewidth}{!}{
  \begin{tabular}{lp{3cm}p{6.2cm}p{4.8cm}}
    \toprule[1.2pt]
    \textbf{Registry Type} & \textbf{Example} & \textbf{Identity and Account} & \textbf{Credential and Trust Info} \\
    \midrule[1.2pt]
    User &
    Human user, org account &
    Account mapped to a unique 32-character identity code; stores basic profile, organizational relationships, and assigned roles in the registry &
    Login secret, certificate, or verifiable credential; encodes user affiliation, role, authorization baseline, and compliance attributes \\ \midrule

    Agent (MCP host) &
    General/vertical agent, orchestrator &
    Each MCP Host assigned a 32-character identity code; registry records code or model hash, deployment environment, functional scope, and operational boundaries &
    Host digital credential with public key, declared capabilities, permission baseline, and optionally delegation or controller information \\ \midrule

    MCP Server &
    Tool services, integration endpoint &
    Each server endpoint or instance issued a unique identity code; registry maintains provider/operator, exposed capability description, and service sensitivity level &
    Service certificate or verifiable credential; includes service identity, supported authentication, compliance and data handling policy, and operational trust attributes \\ \midrule

    Resource &
    Model API, dataset, external service &
    Each resource assigned a unique \texttt{modelId}, \texttt{datasetId}, etc.; registry stores owner, version, provenance, and sensitivity labels &
    Resource-level credential or assertion issued by trust anchor; specifies compliance, permitted usage scope, risk classification, and supports auditability \\
    \bottomrule[1.2pt]
  \end{tabular}
  }
\end{table}

%% file: Tables/context.tex
\begin{table}[htbp]
  \centering
  \caption{Unified Security Context Fields in SMCP.}
  \label{tab:context}
  \resizebox{1\linewidth}{!}{
  \begin{tabular}{p{4cm}lp{9cm}}
    \toprule[1.2pt]
    \textbf{Field} & \textbf{Example Name} & \textbf{Description} \\
    \midrule[1.2pt]
    \multirow{2}{*}{Session Identifier} & \multirow{2}{*}{\texttt{sessionId}} &
    A session ID shared across messages, tasks, and invocations, corresponding to the session concept in agent interaction standards. \\ \midrule

    \multirow{2}{*}{Call Chain Identifier} & \multirow{2}{*}{\texttt{callChainId}} &
    A globally unique ID for a user task, used for end-to-end auditability across agents and tools. \\ \midrule

    \multirow{3}{*}{Delegation Chain} & \multirow{3}{*}{\texttt{delegatorChain}} &
    A summary of the authorization chain from user or organization through the entry agent to the current actor, supporting accountability and policy enforcement. \\ \midrule

    \multirow{3}{*}{Caller Identity} & \multirow{3}{*}{\texttt{callerAgentId}} &
The identity code of the agent initiating the current request, bound to the identity management domain and trusted component registry. \\ \midrule

\multirow{2}{*}{Peer Identity} & \multirow{2}{*}{\texttt{peerId}} &
The identity of the interaction peer (tool service, model, or data service), used for mutual authentication and access control. \\ \midrule

\multirow{2}{*}{Authentication Assertion} & \multirow{2}{*}{\texttt{authnAssertionId}} &
Reference to the most recent authentication assertion, used to prove the current session or channel's assurance level. \\ \midrule

\multirow{3}{*}{Risk Level} & \multirow{3}{*}{\texttt{riskLevel}} &
The risk classification for the current task or invocation (e.g., low, medium, high, critical), driving adaptive authentication and authorization strategies. \\ \midrule

\multirow{2}{*}{Policy Reference} & \multirow{2}{*}{\texttt{policyRef}} &
A pointer to the applicable set of access control and compliance policies, used for decision explanation and auditing. \\ \midrule

\multirow{2}{*}{Data Sensitivity} & \multirow{2}{*}{\texttt{dataSensitivity}} &
The sensitivity level of data involved in this interaction, constraining access to the minimum necessary. \\ \midrule

\multirow{2}{*}{Timestamp Protection} & \multirow{2}{*}{\texttt{timestamp}, \texttt{nonce}} &
Used with underlying TLS or message signatures to resist replay and ordering attacks. \\

    \bottomrule[1.2pt]
  \end{tabular}}
\end{table}

%% file: Tables/object.tex
\begin{table}[htbp]
    \centering
    \caption{Core Objects and Their Roles in SMCP.}
    \resizebox{1\linewidth}{!}{
    \begin{tabular}{llp{8.5cm}}
        \toprule[1.2pt]
        \textbf{Object} & \textbf{Example Identifiers} & \textbf{Description} \\
        \midrule[1.2pt]
        \multirow{2}{*}{Tool Capability Description} & \multirow{2}{*}{\texttt{toolId}, \texttt{sensitivityLevel}} & Uniquely identifies the tool and its security attributes, used for access control and risk assessment before invocation. \\
        \midrule
        \multirow{3}{*}{Tool Invocation Request} & \multirow{3}{*}{\texttt{toolCallId}, \texttt{callChainId}} & Binds a unified security context and component identity references, enabling each invocation to be traced back to upstream sessions, tasks, and underlying components. \\
        \midrule
        \multirow{3}{*}{Tool Execution Response} & \multirow{3}{*}{\texttt{actionCode}, \texttt{auditRef}} & Returns the execution status and result summary, and provides audit record references for the resource access domain and related components. \\
        \bottomrule[1.2pt]
    \end{tabular}}
    \label{tab:object}
\end{table}

%% file: Tables/capacity.tex
\begin{table}[htbp]
    \centering
    \caption{Security Extension Fields for Tool Capabilities (Illustrative).}
    \resizebox{1\linewidth}{!}{
    \begin{tabular}{llp{8.3cm}}
        \toprule[1.2pt]
        \textbf{Field} & \textbf{Example Name} & \textbf{Description} \\
        \midrule[1.2pt]
        \multirow{3}{*}{Sensitivity Level} & 
            \multirow{3}{*}{\texttt{sensitivityLevel}} & 
            Categorizes tools into public, internal, sensitive, or highly sensitive levels, each corresponding to different invocation constraints. \\
        \midrule
        \multirow{3}{*}{Minimum Identity Assurance} & 
            \multirow{3}{*}{\texttt{minIdentityAssurance}} & 
            Requires the invoker to have at least a certain authentication strength (can be mapped to LoA or custom levels in this framework). \\
        \midrule
        \multirow{3}{*}{Allowed Delegators} & 
            \multirow{3}{*}{\texttt{allowedDelegators}} & 
            Specifies which subject types or tenants may access the tool via agent delegation (e.g., only users from the same organization are allowed). \\
        \midrule
        \multirow{4}{*}{Data Policy} & 
            \multirow{4}{*}{\texttt{dataPolicy}} & 
            Indicates whether results may be stored long-term, included in agent long-term memory, used for model retraining, or require redaction; can be linked to compliance labels of underlying resources. \\
        \midrule
        \multirow{3}{*}{Audit Requirement} & 
            \multirow{3}{*}{\texttt{auditRequirement}} & 
            Specifies the minimum granularity for invocation chain records (e.g., whether to record parameter summaries, result summaries, and caller context). \\
        \bottomrule[1.2pt]
    \end{tabular}}
    \label{tab:capacity}
\end{table}

%% file: Tables/field.tex
\begin{table}[htbp]
    \centering
    \caption{Key Security Fields in SMCP Invocation and Response.}
    \resizebox{1\linewidth}{!}{
    \begin{tabular}{l >{\raggedright\arraybackslash}p{3.3cm} p{10.5cm}}
        \toprule[1.2pt]
        \textbf{Direction} & \textbf{Field} & \textbf{Description} \\
        \midrule[1.2pt]
        \multirow{11}{*}{Invocation Request} & 
            \texttt{agentId}, \texttt{authnAssertionId} & 
            Specifies the invoking agent's identity and the current identity assurance status, which can be jointly verified with the tool's \texttt{minIdentityAssurance} requirement. \\
        \cmidrule{2-3}
         & 
            \texttt{delegatorChain}, \texttt{riskLevel}, \texttt{dataSensitivity} & 
            Propagates the delegation chain and risk semantics from upstream sessions to the tool layer, serving as the basis for authorization and downgrade decisions. \\
        \cmidrule{2-3}
         & 
            \texttt{callChainId}, \texttt{sessionId}, \texttt{taskId}, \texttt{toolCallId} & 
            Binds the current tool invocation to the global invocation chain, session, and specific task, forming a traceable invocation segment that can be correlated with tool-side audit records. \\
        \cmidrule{2-3}
         & 
            \texttt{toolId}, \texttt{toolVersion}, \texttt{toolInput} & 
            Describes the capability and parameters compatible with existing tool interfaces to drive the actual execution. \\
        \midrule
        \multirow{6}{*}{Tool Response} & 
            \texttt{actionCode}, \texttt{result} & 
            Indicates the execution status and result data (or summary), compatible with existing invocation patterns. \\
        \cmidrule{2-3}
         & 
            \texttt{timestamp}, \texttt{signature} & 
            Timestamp and integrity protection information signed by the tool service or security gateway, ensuring result non-repudiation. \\
        \cmidrule{2-3}
         & 
            \texttt{auditRef} & 
            Reference to the tool-side local audit record, which can be associated with \texttt{callChainId} and related component identifiers for end-to-end traceability. \\
        \bottomrule[1.2pt]
    \end{tabular}}
    \label{tab:field}
\end{table}

%% file: Tables/log.tex
\begin{table}[htbp]
    \centering
    \caption{Key Attributes Captured in SMCP Audit Log Entries.}
    \resizebox{1\linewidth}{!}{
    \begin{tabular}{l >{\raggedright\arraybackslash}p{4.8cm} p{9.5cm}}
        \toprule[1.2pt]
        \textbf{Attribute Category} & \textbf{Example Fields} & \textbf{Description} \\
        \midrule[1.2pt]
        \multirow{2}{*}{Timestamps} & 
            \texttt{timestamp}, \texttt{signedAt} & 
            Records the time of invocation, response, and log entry signing, supporting chronological reconstruction and non-repudiation. \\
        \midrule
        \multirow{4}{*}{Actor and Subject Identities} & 
            \texttt{agentId}, \texttt{userId}, \texttt{delegatorChain}, \texttt{subjectId}, \texttt{subjectType}, \texttt{role} & 
            Identifies the primary invoker, end user, and any delegation chain; distinguishes between user and agent roles for accountability and compliance tracking. \\
        \midrule
        \multirow{4}{*}{Operation Details} & 
            \texttt{toolId}, \texttt{capabilityRef}, \texttt{actionCode}, \texttt{taskId}, \texttt{sessionId}, \texttt{callChainId}, \texttt{auditRef} & 
            Describes the invoked tool, requested capability, action performed, and unique identifiers for the session, task, and audit chain, enabling granular traceability and correlation. \\
        \midrule
        \multirow{3}{*}{Policy Decisions and Obligations} & 
            \texttt{policyRef}, \texttt{policyDecision}, \texttt{obligations}, \texttt{constraints} & 
            Captures the applicable policy references, decisions (e.g., permit with obligations), imposed constraints, and obligations such as data masking or aggregation. \\
        \midrule
        \multirow{2}{*}{Sensitivity and Risk Level} & 
            \texttt{requestedSensitivityLevel}, \texttt{appliedSensitivityLevel} & 
            Indicates the sensitivity and risk classification requested and actually applied to the data, supporting risk-aware auditing and compliance. \\
        \midrule
        \multirow{2}{*}{Result Metadata} & 
            \texttt{resultSummary}, \texttt{metrics}, \texttt{resultDetailsRef} & 
            Summarizes the outcome of the operation, including key performance indicators and references to detailed result data. \\
        \midrule
        \multirow{2}{*}{Signatures and Integrity} & 
            \texttt{signature}, \texttt{hashAlg}, \texttt{signedBy} & 
            Provides cryptographic signatures, hash algorithms, and signer identity to ensure log integrity and enable tamper evidence. \\
        \bottomrule[1.2pt]
    \end{tabular}}
    \label{tab:log}
\end{table}

%% file: Sections/5.Usecase.tex
\section{SMCP Use Cases}
\label{sec:usecase}

To illustrate the practical application and security features of the SMCP, we present a representative use case that demonstrates its complete workflow. This example describes how the main mechanisms of SMCP, including secure authentication, policy enforcement, capability-based access control, and audit logging, cooperate to address security challenges in a modular software system.

As outlined in \autoref{tab:usecase}, the process begins with the user and tools performing mutual authentication through the SMCP authentication module. This initial verification ensures that only trusted participants are allowed to proceed. Following authentication, the user searches for available tools by querying the centralized registry. The registry applies strict naming policies and verifies metadata, which helps to prevent issues such as typosquatting and the registration of unauthorized or potentially malicious tools.
Once a suitable tool is identified, the user submits a request to invoke the tool. Before the request is processed, the SMCP policy engine evaluates it according to predefined security and capability policies. This step ensures that unauthorized or unsafe operations are blocked before execution. During invocation, each tool operates within a restricted security context determined by SMCP, thereby minimizing privilege exposure and limiting access to sensitive resources.

\input{Tables/usecase}

%% file: Tables/usecase.tex
\begin{table}[htbp]
    \centering
    \caption{Mapping of Security Risks to SMCP Mitigation Steps.}
    \resizebox{\linewidth}{!}{
    \begin{tabular}{llp{10.5cm}}
        \toprule[1.2pt]
        \textbf{Risk Type} & \textbf{SMCP} & \textbf{Mitigation Details} \\
        \midrule[1.2pt]
        Unauthenticated Access & \autoref{subsec:authentication} & Enforces strong mutual authentication and secure exchange of credentials to prevent unauthorized entities from accessing any part of the system. \\
        \midrule
        Session Management Flaws & \autoref{subsec:authentication} & Secures session initiation, renewal, and termination to prevent session hijacking and unauthorized reuse. Ensures active session validation throughout the workflow. \\
        \midrule
        Namespace Typosquatting & \autoref{subsec:registry} & Trusted registry enforces strict naming policies and validation, blocking deceptively named or malicious components from being registered or discovered. \\
        \midrule
        Tool Name Conflict & \autoref{subsec:registry} & Registry ensures tool identifier uniqueness with conflict checks and prevents ambiguity or accidental invocation of the wrong tool. \\
        \midrule
        Tool Poisoning & \autoref{subsec:registry}, \autoref{subsec:policy} & Only verified tools are registered; policy engine validates tool metadata and enforces behavior checks before invocation, reducing risk of malicious tool insertion. \\
        \midrule
        Command Injection & \autoref{subsec:policy} & Policy rules and input validation filter unsafe payloads, preventing malicious command execution in tool calls or actions. \\
        
        \bottomrule[1.2pt]
    \end{tabular}}
    \caption*{\hfill(Continued)}
    \label{tab:usecase}
\end{table}

\begin{table}[htbp]
    \captionsetup{labelformat=empty}
    \caption*{Table \ref{tab:usecase}. Continued.}
    \centering
    \resizebox{\textwidth}{!}{
    \begin{tabular}{llp{10.5cm}}
        \toprule[1.2pt]
        \textbf{Risk Type} & \textbf{SMCP} & \textbf{Mitigation Details} \\
        \midrule[1.2pt]      
        Rug Pulls & \autoref{subsec:registry} & Registry maintains version history and provenance for all tool entries, preventing undetected malicious updates or sudden removals. \\
        \midrule
        Cross-Server Shadowing & \autoref{subsec:registry} & Centralized registry ensures authoritative mapping and prevents registration of duplicate or shadow tool endpoints. \\
        \midrule
        Prompt Injection & \autoref{subsec:context}, \autoref{subsec:policy} & Security context constrains available agent actions and policy checks ensure prompt contents cannot escalate privileges or induce undesired behaviors. \\
        \midrule
        Indirect Prompt Injection & \autoref{subsec:context}, \autoref{subsec:policy} & Policy checks and secure context propagation restrict the impact of adversarial manipulation in indirect or multi-stage prompts. \\
        \midrule
        Installer Spoofing & \autoref{subsec:registry} & Authenticated, integrity-checked distribution of installers and updates blocks use of tampered or fake installation sources. \\
        \midrule
        Preference Manipulation & \autoref{subsec:registry} & Registry verifies tool descriptions and metadata are signed and authentic, protecting against attacker-influenced preference modifications. \\
        \midrule
        Remote Code Execution & \autoref{subsec:policy} & All inputs are validated and policy controls restrict code execution scope, minimizing the risk from untrusted or injected code. \\
        \midrule
        SQL Injection & \autoref{subsec:policy} & Policy engine includes input validation for data operations, effectively blocking injection attacks in tool or server interactions. \\
        \midrule
        Credential Theft & \autoref{subsec:context}, \autoref{subsec:policy} & Security context traces credential use; least-privilege access design and scoping minimize possible credential exposure or misuse. \\
        \midrule
        Sandbox Escape & \autoref{subsec:policy} & Policy and runtime isolation controls restrict the execution environment of tools, limiting their ability to break out of sandboxes. \\
        \midrule
        Tool Chaining Abuse & \autoref{subsec:policy} & Capability models and policy logic limit tool chaining depth and validate all chained calls to prevent escalation or abuse. \\
        \midrule
        Unauthorized Access & \autoref{subsec:policy} & Fine-grained policy and capability enforcement ensure no component receives excess permissions, reducing attack surface. \\
        \midrule
        Privilege Abuse & \autoref{subsec:policy}, \autoref{subsec:log} & Policy restricts privilege escalation; audit logs provide full traceability and enable detection of abnormal privilege use. \\
        \midrule
        Path Traversal & \autoref{subsec:policy} & Policy checks enforce strict path validation for all resource accesses, blocking unauthorized or dangerous file operations. \\
        \midrule
        Resource Content Poisoning & \autoref{subsec:policy}, \autoref{subsec:log} & Policy validation and comprehensive audit logging enable early detection of manipulated or malicious resource results. \\  
        \midrule
        Token Passthrough & \autoref{subsec:context}, \autoref{subsec:log} & Scoped context design restricts token propagation; audit logs detect and alert on any inadvertent credential forwarding. \\
        \midrule
        Cross-Tenant Data Exposure & \autoref{subsec:context}, \autoref{subsec:log} & Security context enforces strict tenant isolation; audit logs track and support investigation of any data leakage events. \\
        \midrule
        Vulnerable Versions & \autoref{subsec:registry}, \autoref{subsec:log} & Registry enforces component version control; audit logs help identify outdated or unpatched components for timely updates. \\
        \midrule
        Privilege Persistence & \autoref{subsec:log} & Audit trail tracks every privilege modification, supporting detection and remediation of improper retention or persistence of access. \\
        \midrule
        Configuration Drift & \autoref{subsec:registry}, \autoref{subsec:log} & Registry maintains a reference configuration baseline; auditing reveals unauthorized or accidental configuration changes. \\
        \bottomrule[1.2pt]
    \end{tabular}}
\end{table}

%% file: Sections/6.Relatedwork.tex
\section{Related Work}
\label{sec:related}

\subsection{MCP Security}
Since the release of the MCP, the security community has rapidly moved to identify potential vulnerabilities inherent in its open and modular design. Early efforts have focused on cataloging risks and establishing benchmarks for attack surfaces. \texttt{Adversa.ai} released a comprehensive list of the top 25 MCP vulnerabilities, highlighting issues ranging from unauthorized access to supply chain risks~\cite{mcptop25}. Meanwhile, academic researchers have sought to formalize these risks within the protocol's architecture. Hou et al. provided a comprehensive landscape analysis of MCP, detailing security threats and future research directions~\cite{hou2025mcp}. In parallel, Guo et al. conducted a systematic analysis of MCP security, introducing MCPLib to categorize threats across the client-host-server interaction model~\cite{MCPLIB}.
Building on these foundational taxonomies, recent work has exposed specific attack vectors arising from the trust agents place in external servers. Zhao et al. treated servers as active threat actors, identifying twelve distinct attack categories where malicious implementations can exploit host-LLM interactions~\cite{zhao2025mcpserversattacktaxonomy}. Specific vulnerabilities in the registration phase were highlighted by Wang et al. via MCPTox, which demonstrated how tool poisoning can embed malicious instructions in metadata to manipulate agent behavior~\cite{MCPTox}. Beyond direct attacks, the complexity of toolchains introduces indirect risks; Zhao et al. revealed parasitic toolchain attacks, where passive data sources inject prompts to hijack execution flows without direct user interaction~\cite{zhao2025mindyourserver}. Furthermore, Wang et al. uncovered economic threats through MPMA, showing how servers can manipulate descriptions to bias LLM selection for financial gain~\cite{wang2025mpma}.
In response to these threats, the community has moved towards empirical assessment and defensive frameworks. Hasan et al. conducted a large-scale empirical study of open-source MCP servers, finding that despite high community engagement, many servers suffer from maintainability issues and vulnerabilities like credential exposure~\cite{hasan2025mcpglance}. To enable proactive security, Radosevich and Halloran developed McpSafetyScanner, an agentic tool for auditing servers against exploits such as remote access control before deployment~\cite{radosevich2025mcpsafetyauditllms}. On the benchmarking front, Yang et al. introduced MCPSecBench to systematically evaluate 17 attack types across the protocol's layers, exposing weaknesses in current host defenses~\cite{yang2025mcpsecbench}. Finally, Shi et al. proposed SecMCP, a defense mechanism that leverages latent polytope analysis to detect conversation drift, effectively identifying adversarial deviations in agent trajectories~\cite{shi2025quantifyingconversationdriftmcp}.

\subsection{Agentic AI System Security}

Parallel to protocol-specific research, the broader field of agentic AI security has developed various defense mechanisms focusing on the agent's runtime environment and decision-making processes. Comprehensive surveys by Gan et al.~\cite{gan2023maas} and Li et al.~\cite{Li2024Personal} have categorized threats facing LLM-based agents, emphasizing the need for robust defenses against prompt injection and unsafe action execution.
To mitigate these risks, researchers have moved beyond static rules towards dynamic runtime protection. AgentArmor utilizes program analysis to trace agent behavior~\cite{AgentArmor}, while AgentSpec enforces formal contracts to prevent deviations~\cite{AgentSpec}. Advancing this dynamic capability, Xiang et al. proposed GuardAgent, which employs a dedicated agent to interpret safety requests and generate executable guardrail code via knowledge-enabled reasoning, ensuring compliance without altering the target agent's core logic~\cite{xiang2025guardagentsafeguardllmagents}. Shifting from reactive to proactive defense, Wang et al. introduced PRO$^2$GUARD, a framework that leverages discrete-time Markov chains to probabilistically predict unsafe agent trajectories, enabling interventions before violations actually occur~\cite{wang2026pro2guardproactiveruntimeenforcement}.
In the domain of interaction integrity and access control, recent works have focused on granular policy enforcement. Jing et al. developed the Model Contextual Integrity Protocol (MCIP), which integrates a guardian model and tracking tools to enforce information flow integrity based on a fine-grained risk taxonomy~\cite{jing-etal-2025-mcip}. Complementing this, Shi et al. presented Progent, a programmable privilege control framework that enforces the principle of least privilege by intercepting tool calls with deterministic security policies, effectively reducing the attack surface for coding and web agents~\cite{shi2025progentprogrammableprivilegecontrol}. These approaches sit alongside architectures like SAGA~\cite{SAGA} and isolation techniques such as DRIFT~\cite{DRIFT} and A2AS~\cite{neelou2025a2asagenticairuntime}, creating a multi-layered defense landscape.


%% file: Sections/7.Conclusion.tex
\section{SMCP Roadmap}
\label{sec:roadmap}

The SMCP is being developed as an open and extensible security foundation, providing unified digital identity, contextual access control, and end-to-end auditability for agent-tool interactions.

\noindent\textbf{NEAR TERM}, the goal is to integrate SMCP’s core security mechanisms, such as trusted component registries, structured digital identity codes, and unified security context propagation, into existing MCP implementations and widely-used agent frameworks. This stage will focus on developing reference implementations, facilitating compatibility with current agent architectures, and demonstrating feasibility in real-world engineering deployments. Additional research will target best practices for entity registration workflows, authentication procedures, and security context message formats.

\noindent\textbf{MID TERM}, the focus will shift to expanding SMCP’s policy enforcement and runtime auditing capabilities. Key directions include developing fine-grained risk-adaptive access control engines, scalable audit logging infrastructure, and advanced delegation and revocation management. Industry feedback and empirical studies will inform improvements in policy expressiveness, compliance automation, and operational efficiency. This phase will also explore interoperability with other agent security standards and protocols.

\noindent\textbf{LONG TERM}, the roadmap envisions SMCP as a foundational layer for secure, interoperable agent ecosystems across diverse domains and trust boundaries. Future work will include supporting multi-domain federation, automated trust negotiation, and integration with evolving AI-native protocols and governance frameworks. Ultimately, the goal is to establish SMCP as a reference standard for secure agent interconnection and collaborative AI governance, enabling scalable, accountable, and trustworthy agentic AI on a global scale.

\section{Conclusion}
\label{sec:conclusion}

This paper presented the SMCP, a comprehensive security framework designed to address the unique challenges of open, MCP-based agent ecosystems. We systematically analyzed the security risks inherent in the MCP workflow, which arise from the protocol’s open and composable architecture. While existing research has made significant progress in attack analysis and agent-level defenses, there remains a critical gap in protocol-native, end-to-end security enhancements for MCP systems. SMCP fills this gap by introducing unified digital identity and trust infrastructure, mutual authentication, continuous security context propagation, fine-grained policy enforcement, and structured audit logging directly at the protocol layer. Through these mechanisms, SMCP delivers robust access control, adaptive runtime protection, and comprehensive accountability across agent-tool interactions. Our analysis and use cases demonstrate that SMCP can effectively mitigate the most pressing security threats in modern agentic workflows while maintaining flexibility and interoperability. Looking ahead, the SMCP roadmap envisions its integration into mainstream agent architectures and standardization as a foundational layer for secure, scalable, and trustworthy agentic AI ecosystems.